\documentclass[a4paper,english,oneside,11pt,final]{book}

\usepackage[utf8]{inputenc}
\usepackage[T1]{fontenc}
\usepackage[italian,english]{babel}

\usepackage{lmodern,textcomp}
\usepackage[final]{easyoutput}
\usepackage{makeidx} % Indice analitico
	\makeindex

\usepackage{natbib}
	\setcitestyle{numbers,author,square}
\usepackage[binding=0.8cm]{layaureo}  % FOR PRINTABLE VERSION: migliore copertura foglio A4 + margine rilegatura
\usepackage{tocbibind} 
\usepackage[labelfont=bf]{caption}% didascalie figure
\usepackage{subfig}

\usepackage[cmex10]{amsmath}
	\mathchardef\minus="002D
\usepackage{amssymb,amsfonts,amsthm,mathrsfs}
	\theoremstyle{plain}
	\newtheorem*{theorem*}{Theorem}
	\newtheorem*{weakdef*}{Weak equivalence principle}
	\newtheorem*{strongdef*}{Strong equivalence principle}
	\newtheorem*{massdef*}{Mass superselection rule}
\usepackage{physics}

\usepackage{booktabs}
\usepackage{tabulary}

\usepackage[disable]{todonotes}

\usepackage{verbatim}

\usepackage{cleveref}
	\crefname{section}{§}{§§}
	\Crefname{section}{§}{§§}

\usepackage[smaller]{acronym}
	\acrodef{QM}{quantum mechanics}
	\acrodef{CQM}{classical quantum mechanics}
	\acrodef{EGT}{extended Galilean transformation}
	\acrodef{QW}{quantum walk}

	\acused{CQM}
	\acused{QW}

\newcommand{\keyword}[1]{\emph{#1}\index{#1}}
\newcommand{\DeltaVar}[1]{\Delta #1\,}

\newcommand{\Bargmann}{V.~Bargmann}
\newcommand{\Greenberger}{D.~M.~Greenberger}
\newcommand{\DanielGreenberger}{Daniel M.~Greenberger}
\newcommand{\ie}{i.~e.}
\newcommand{\eg}{e.~g.}

\title{Mass and Proper Time as Conjugated Observables}
\author{Matteo Lugli}
\date{Anno accademico 2016/2017}

\begin{document}\selectlanguage{italian}
\frenchspacing
\frontmatter
%%%%%%%%%%%%%%%%%%%%%%%%%%%%%%%%%%%%%%%%%%%%%%%%%%%%%%%%%%%%
% Frontespizio
%%%%%%%%%%%%%%%%%%%%%%%%%%%%%%%%%%%%%%%%%%%%%%%%%%%%%%%%%%%%
\begin{titlepage}
	\begin{center}
		\vskip 1cm 
		\includegraphics[width=4cm]{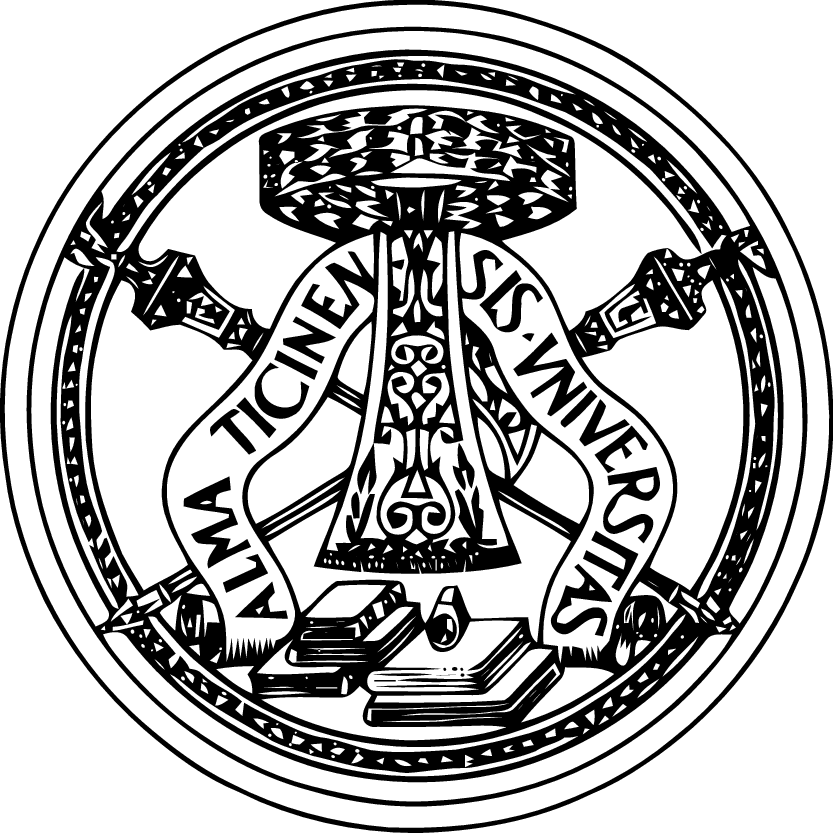}
		\vskip 0.5cm 
		
		\LARGE
			\textbf{Università degli Studi di Pavia}\\
			\textbf{Dipartimento di Fisica}\\
			\vskip 0.5cm 
		\Large
			Corso di Laurea Triennale in Fisica
		
		\vskip 1.5cm
		\Huge
			\textbf{\foreignlanguage{english}{\printtitle}}
		\vskip 2.5cm
	\end{center}

	\Large
	\begin{minipage}[b]{10.5cm}
		{\large Relatore:}\\
		\textsc{Chiar.mo prof. Giacomo Mauro d'Ariano}\\
		\\
		{\large Correlatore:}\\
		\textsc{Dott. Alessandro Tosini}
	\end{minipage}
	\vskip 1cm
		
	\hfill
	\begin{minipage}[t]{4cm}
		{\large Tesi di laurea di:}\\
		\textsc{\printauthor}\\
		Matricola: 431102
	\end{minipage}
	
	\vfill
	
	\begin{center}
		\printdate
	\end{center}
	
	\eject
\end{titlepage}

%%%%%%%%%%%%%%%%%%%%%%%%%%%%%%%%%%%%%%%%%%%%%%%%%%%%%%%%%%%%
% Dedica
%%%%%%%%%%%%%%%%%%%%%%%%%%%%%%%%%%%%%%%%%%%%%%%%%%%%%%%%%%%%

\cleardoublepage

\vspace*{10pc}
\begin{flushright}
	\sl
	
	Alla mia famiglia	
\end{flushright}

%\leavevmode\thispagestyle{empty}\newpage
%
%\vspace*{10pc}
%\thispagestyle{empty}
%\begin{flushright}
%\sl
%
%Alla mia famiglia \\
%
%\end{flushright}
%
%\par\vfill\par

%%%%%%%%%%%%%%%%%%%%%%%%%%%%%%%%%%%%%%%%%%%%%%%%%%%%%%%%%%%%
% Indici
%%%%%%%%%%%%%%%%%%%%%%%%%%%%%%%%%%%%%%%%%%%%%%%%%%%%%%%%%%%%
\selectlanguage{english}
\tableofcontents

\mainmatter
%%%%%%%%%%%%%%%%%%%%%%%%%%%%%%%%%%%%%%%%%%%%%%%%%%%%%%%%%%%%
%\chapter*{Preface}
%%%%%%%%%%%%%%%%%%%%%%%%%%%%%%%%%%%%%%%%%%%%%%%%%%%%%%%%%%%%
%\label{cap:pre}

%\addcontentsline{toc}{chapter}{Preface}

%%%%%%%%%%%%%%%%%%%%%%%%%%%%%%%%%%%%%%%%%%%%%%%%%%%%%%%%%%%%
\chapter{Introduction}
%%%%%%%%%%%%%%%%%%%%%%%%%%%%%%%%%%%%%%%%%%%%%%%%%%%%%%%%%%%%

This work discloses some inconsistencies of \ac{QM} in two different contexts.
On the one hand the equivalence principle in his two formulations is the starting point of all gravitational theories; yet there is no way to make it consistent with the non-relativistic formulation of \ac{QM}.
Moreover, the evaluation of the Schrödinger equation for a particle in an \emph{accelerated} reference frame sheds light on a trajectory-dependent phase term subject to proper time.
On the other hand, the Bargmann theorem unmistakably proves that the Schrödinger equation does not admit any superposition of different masses if the symmetry chosen for the system is the Galilean one.
However, this is incompatible with relativity, since mass and energy are equivalent and one can certainly superimpose different energies.
Both inconsistencies support the hypothesis that mass and proper time could be treated as variables.
In the third chapter we will describe a theory in which the Galilean symmetry is extended by assuming particle mass and proper time as canonically conjugated classical variables, or conjugated observables in the quantum theory.

The first three chapters discuss subjects illustrated in several articles of the professor \DanielGreenberger{}, which are listed in the bibliography.
Whereas in the last chapter the results of a recent theory on \aclp{QW}, say a discrete model for quantum particle evolution, will be examined.
In particular, the striking result is that, in the most natural physical interpretation of \acl{QW} dynamics, it is not possible to have a constant and invariant mass in any manner.
The long term aim of this work is indeed to study the reasons and consequences of a theory of particles with variable mass in the context of wider studies still in progress on \aclp{QW}.

%\vfill
%\begin{description}
%	\item[\DanielGreenberger] (born 1932) is a physics professor at the City College of New York located in New York NY, United States.
%	He is also a fellow of the American Physical Society. \\
%	Web page: \url{https://www.ccny.cuny.edu/profiles/daniel-greenberger}
%\end{description}

%%%%%%%%%%%%%%%%%%%%%%%%%%%%%%%%%%%%%%%%%%%%%%%%%%%%%%%%%%%%
\chapter{The Equivalence Principle in Quantum Mechanics}
\chaptermark{The Equivalence Principle in \ac{QM}}
%%%%%%%%%%%%%%%%%%%%%%%%%%%%%%%%%%%%%%%%%%%%%%%%%%%%%%%%%%%%
\label{cap:time}\acresetall

In this chapter we will discuss the equivalence principle both in classical and \ac{QM}.
Two different formulations of the principle are introduced and their incompatibilities with the quantum postulates.

The very first version of the principle has been studied for centuries by physicists, yet it reveals the most profound and insurmountable inconsistencies with the contemporary theory of \acl{QM}.
On the other hand, the latest version introduced by A.~Einstein, which is the cornerstone of general relativity, has a more elaborated and modern approach. Nevertheless, the assumption of this principle in \ac{QM} leads to subtle results, whose physical effects have not yet been observed.
None of the definitions can actually be applied in \acl{QM}, partly because the classical conditions built in the formulations have no equivalent in \ac{QM}.

Finally, we will see how the curvature of spacetime due to the gravitational field, in a general relativity context, is able to legitimize the upgrade of proper time from a kinematical to a dynamical.

The subjects of this chapter are treated in \cite{Greenberger1968,Greenberger1970-I,Greenberger1970-II}, and in the paper \cite{Greenberger2010b}.

%***********************************************************
\section{Weak and strong equivalence principles}
%***********************************************************
\label{sec:weakandstrong}

Let us consider a point-like non-relativistic massive particle immersed in an \emph{external gravitational field}\index{approximation!external gravitational field}. In this approximation, the gravitational field is generated by a much more massive distribution and therefore the particle perturbation to the field is negligible.
In such conditions, the weak or Galilean equivalence principle, also known as universality of free fall, states:

\begin{weakdef*}\index{equivalence principle!weak}
	The inertial mass $m_i$, \ie{} the physical quantity measured with a balance, and weight or gravitational mass $m_g$, the physical quantity measured with a scale, are locally in identical ratio for all bodies. Henceforth we will assume:
	\begin{equation}
		m_i = m_g .
	\end{equation}
\end{weakdef*}

Letting $m = m_i = m_g > 0$, we can set up in Hamiltonian formalism the equations of motion for the particle:
\begin{equation}
\begin{split}
H(q,p,t) = \frac{p^2}{2m_i} + m_gV_G(q) = \frac{p^2}{2m} + mV_G(q), \quad \{q,p\} = 1 \\
\dot{q} = + \pdv{H}{p}, \quad \dot{p} = - \pdv{H}{q}
\end{split}
\end{equation}
where the over-dot $\dot{x} = \dv{x}{t}$  is time derivative in Newton's notation, $\pb{f}{g}$ are the Poisson brackets, and $V_G$ is the gravitation potential.
To give an example, for a small particle orbiting close to a mass $M$ with $M \gg m$ (external field approximation) in the center-of-mass frame we have:
\begin{equation}\label{eq:gravPot}
V_G(r) = G \frac{M}{r}
\end{equation}
with $G$ the universal gravitational constant.

Now, defining the velocity varible and a new Hamiltonian function as follows:
\begin{equation}\label{eq:masslessQuantities}
v = \frac{1}{m} p, \quad \mathcal{H} = \frac{H}{m} = \frac{1}{2} v^2 + V_G(q)
\end{equation}
we obtain a formalism totally independent of the mass $m$ and with the same solutions of the previous one.
This can be proved with ease by considering the new Poisson brackets and Hamilton equations:
\begin{equation}
\begin{split}
\pb{f}{g}_{x,v} = \pdv{f}{x}\pdv{g}{v} - \pdv{g}{x}\pdv{f}{v}, \quad\pb{x}{v}_{x,v} = 1 \\
\dot{q} = + \pdv{H}{v}, \quad\dot{v} = - \pdv{H}{q} .
\end{split}
\end{equation}
Thus, it is elementary to show that the following formulation of the principle is equivalent to the above one:

\begin{weakdef*}
	The vacuum world-line of a body immersed in a gravitational field is independent of all observable properties.
\end{weakdef*}

As a consequence, if two point particles are released with the same initial conditions (position and velocity) in the same gravitational field, they will travel along the same trajectories regardless of their masses.
The latter statement provides a geometrical interpretation of the (weak) equivalence principle and therefore it is able to show how fundamental it is.
For this reason one might expects these statements to hold in all mechanics theories.

As soon as one tries to verify the validity of the statements in \acl{QM}, they notice how inapplicable the second formulation is, since no trajectories are available in \ac{QM} to the slightest degree.
Accordingly, we shall see whether the kinematical quantities of motion are at least independent of the mass.

The first inconsistency can be obtained for a \emph{free} particle by recalling the Heisenberg uncertainty relation\index{uncertainty relation!Heisenberg}:
\begin{equation}\label{eq:Heisenberg}
\DeltaVar{x}\DeltaVar{p} \geq \frac{\hbar}{2} . \quad\footnote{$\Delta x$ is formally defined as the standard deviation of the $x$ operator evaluated on the considered state, \ie{} $\Delta x = \sqrt{\sigma_x} = \sqrt{\expval{x^2} - \expval{x}^2}$. A similar definition is available for the momentum $p$.}
\end{equation}
As (\ref{eq:Heisenberg}) includes the particle's position and linear momentum (which contains itself the mass), by preparing a minimally uncertainty initial wave packet for which:
\begin{equation*}
	\DeltaVar{x}\DeltaVar{p} \approx\hbar \rightarrow \DeltaVar{x} \approx\frac{\hbar}{\DeltaVar{p}}\approx\frac{\hbar}{m \DeltaVar{v}} = \frac{\hbar}{m}\frac{1}{\DeltaVar{v}}
\end{equation*}
one is able to calculate the particle's mass from position and velocity spread, which violates the principle of equivalence.

Another instance for a \emph{bound} particle is the gravitational Bohr atom, where one replaces the Coulomb potential with the gravitational one of \cref{eq:gravPot}.
For $M \gg m$, in the center-of-mass frame ($\mu \cong m$ \footnote{The reduced mass $\mu$ is the ``effective'' inertial mass appearing in the two-body problem. It is defined as $\mu = \frac{m M}{m + M}\approx m$ if $M \gg m$.}) and in external field approximation one can solve the new Schrödinger equation by directly replacing the coupling constant (in CGS\footnote{In detail, the electrostatic centimetre–gram–second system of units or ESU-CGS.}):
\begin{equation*}
	e^2 \mapsto G M m .
\end{equation*}

\begin{table}[b]
	\centering
	\caption{Expectation values for the gravitational Bohr atom eigenstates $\psi_{nlm}$ with $l = n$.}
	\label{tab:gravBohrAtom}
	\begin{tabular}{||l|ll||}
		\hline
		Physical quantity &	Bohr atom	& Gravitational Bohr atom \\
		\hline
		\rule{-4pt}{20pt}
		Potential:
		&	$ V_C (r) = \dfrac{e^2}{r} $	&  $\rightarrow V_G (r) = G \dfrac{M m}{r}$ \\ [2ex]
		\rule{-4pt}{20pt}
		Radius:
		&	$\expval{r}_n = \dfrac{\hbar^2}{m e^2} n \left(n - \frac{1}{2}\right)$
		&	$\rightarrow\expval{r}_n = \dfrac{n \left(n - \frac{1}{2}\right)}{G M}\dfrac{\hbar^2}{m^2}$ \\ [2ex]
		\rule{-4pt}{20pt}
		Angular velocity:
		&	$\expval{\omega}_n \simeq\dfrac{n \hbar}{m \expval{r^2}_n}$
		&	$\rightarrow\expval{\omega}_n\simeq \dfrac{G^2 M^2}{n \left(n^2 - \frac{3}{2} n + \frac{1}{2}\right)} \dfrac{m^3}{\hbar^3} $ \\ [2ex]
		\rule{-4pt}{20pt}
		Energy:
		&	$E_n = - \dfrac{m e^4}{2 \hbar^2 n^2}$	&	$\rightarrow E_n = - G^2 \dfrac{M^2}{2 n^2}\dfrac{m^3}{\hbar^2}$ \\ [2ex]
		\hline
	\end{tabular}
\end{table}

The most relevant results are shown in~\cref{tab:gravBohrAtom}.
The total angular momentum is set to $l = n$, so that $\expval{\omega}_n \neq 0$, .
As in \cite[Table~1.4]{Carretta2015}:
\begin{itemize}
\item $\expval{r}_{nl} = \frac{a_0}{2}\left[3 n^2 - l(l + 1)\right]\stackrel{l = n}{=} a_0 n \left(n - \frac{1}{2}\right)$
\item $\expval{r^2}_{nl} = \frac{n^2}{2} a_0^2 \left[5n^2 + 1 - 3l(l + 1)\right] \stackrel{l = n}{=} \frac{1}{2} n^2 a_0^2\left[2 n^2 - 3 n + 1\right]$
\end{itemize}
where $a_0 = \frac{\hbar^2}{m_e e^2}$ is the Bohr radius.
From $L = m \omega r^2 = l\hbar$, one comes to:
\begin{equation*}
	\expval{\omega}_n \stackrel{l = n}{=} \expval{\frac{n \hbar}{m r^2}}_n \simeq\frac{n\hbar}{m\expval{r^2}_n}
\end{equation*}

We can see here, that by measuring \eg{} the average radius of the particle, one can estimate its mass $m$; again, the principle of equivalence is violated in \acl{QM}.
Moreover, all expectation values related to the ``trajectory'' (excluded the energy), are functions of a power of $\frac{\hbar}{m}$.

We remark that the expressions in~\cref{tab:gravBohrAtom} are partly different from those of \cite{Greenberger1968,Greenberger1970-II}.
The reason is that \Greenberger{} uses the Bohr-Sommerfeld quantization rule:
\begin{equation}\label{eq:BohrSommerfeld}
\oint p \dd{q} = n h
\end{equation}
and assumes that electrons follow classical circular orbits.
Instead, the formulae of this work are derived from the solutions to the Schrödinger equation of the Bohr atom.

In order to understand why all this inconsistencies occur, we shall look more deeply at the procedure that led to a classical consistent formulation of the equivalence principle and ask why this fails in the quantum scenario.
On one side we showed that through the substitution in \cref{eq:masslessQuantities} one gets a classical theory that is mass independent.
However, as presented in the following, the mass dependence is reintroduced in the quantum case via the quantization rules.

Hence, by defining the \keyword{velocity operator} as a rescaling of the momentum one:
\begin{equation*}
	\widehat{v} = \frac{1}{m} \widehat{p}
\end{equation*}
where the notation $\widehat{p}$ empathizes the relation being between operators, we infer the new commutation rule:
\begin{equation*}
	\comm{x}{p} = \comm{x}{mv} = i \hbar \longrightarrow \comm{x}{v} = i \frac{\hbar}{m} .
\end{equation*}

By attempting to remove the mass from both the Hamiltonian and the expectation values, we get a \emph{mass-dependent} canonical quantization\index{canonical quantization}, whose procedure quantizes the dynamics, not the kinematics, of the problem.
As a consequence, for a particle in central gravitational force field, the mass appears in all the $\expval{r^n}$ expectation values as a power of $\frac{\hbar}{m}$.

Furthermore, from our construction of (\ref{eq:masslessQuantities}) it is intuitive to see that, for a classical particle in a gravitational field, the energy is proportional to the mass $m$, as $H = m \mathcal{H}$ and $\mathcal{H}$ is independent of the mass.
For two different particles with the same initial condition in the same gravitational field, their binding energies are proportional to their masses, since $\mathcal{H}$ is the same for both and they follow identical trajectories.
However, the $\psi_{nlm}$ state energy of the gravitational Bohr atom, namely $E_n$ on the right side of~\cref{tab:gravBohrAtom}, is proportional to the third power of the mass:
\begin{equation*}
	E_n \propto m^3
\end{equation*}
and hence, it is inconsistent.

As regards the strong equivalence principle, the evaluation of its consistency with the \acl{QM} is quite more straightforward.
Formulated by A.~Einstein first in \cite{Einstein1908,Einstein1911}, it states:

\begin{strongdef*}\index{equivalence principle!strong}
	All effects of a uniform gravitational field are identical to the effects of a uniform acceleration of the coordinate system.
\end{strongdef*}

A similar idea could be retrieved in the classical Newtonian gravitation theory, in which a uniform acceleration of the coordinate system gives rise to a supplementary uniform gravitational field.
\begin{equation*}
	F = ma \stackrel{a' = a + g}{\longmapsto} F - mg = ma' .
\end{equation*}

However, the Einstein equivalence principle generalizes the idea to all laws of physics, including Maxwell's equations of electromagnetism.
For further reading see \cite[p.~190]{Misner1973}.
 
Hence, let us consider a coordinate transformation from the observer inertial reference frame $\left(x,t\right)$ to a uniform accelerated one $\left(x',t'\right)$.
The Galilean transformation is inadequate as it boosts the system to a frame moving with constant velocity, therefore we will use instead the \acfi{EGT}\index{extended Galilean transformation}\index{EGT|see {extended Galilean transformation}}, expressed as follows:
\begin{equation}\label{eq:EGT}
	x' = x + \xi (t), \quad t' = t
\end{equation}
where $\xi\in\mathcal{C}^2$ at least.
Given that every point in space in the frame is accelerating at the same rate, one can see that the new coordinate system is still rigid.

\begin{table}[b]
	\centering
	\caption{Extended Galilean transformation rules for space and time derivatives.}
	\label{tab:EGT}
	\begin{tabular}{||rl||}
		\rule{-4pt}{20pt}
		$\displaystyle\pdv{x}\longrightarrow$ & $\displaystyle\pdv{x'}{x}\pdv{x'} + \pdv{t'}{x}\pdv{t'} = \pdv{x'}$ \\ [.4cm]
		$\displaystyle\pdv{t}\longrightarrow$ & $\displaystyle\pdv{t'}{t}\pdv{t'} + \pdv{x'}{t}\pdv{x'} = \pdv{t'} + \dot{\xi}\left(t\right)\pdv{x'}$ \\ [.4cm]
	\end{tabular}
\end{table}

The wave function $\psi$ for a free particle of mass $m$ is a solution of the Schrödinger equation:
\begin{equation}\label{eq:Schroedinger}
	-\frac{\hbar^2}{2m}\pdv[2]{\psi}{x} = i\hbar\pdv{\psi}{t}
\end{equation}
and by applying the transformations of (\ref{eq:EGT}) and~\cref{tab:EGT} one comes to:
\begin{equation}\label{eq:SchroedingerEGT}
	-\frac{\hbar^2}{2m}\pdv[2]{\psi}{{x'}} = i\hbar\left(\pdv{\psi}{t'} + \dot{\xi}\pdv{\psi}{x'}\right) .
\end{equation}

In order to remove the momentum-dependent term on the right hand side of the equation we introduce a phase factor into the wave function:
\begin{equation}\label{eq:phase}
	\psi\left(x',t'\right) = e^{if\left(x',t'	\right)}\varphi\left(x',t'\right)
\end{equation}
and we evaluate the derivatives in~\cref{eq:SchroedingerEGT}:
\begin{subequations}\label{eq:derivatives}
\begin{alignat}{3}
	\pdv{x'}e^{if}\varphi &= \left(i\pdv{f}{x'}\varphi + \pdv{\varphi}{x'}\right)e^{if}\\
	\pdv[2]{{x'}} e^{if}\varphi &= \left[-\varphi\left(\pdv{f}{x'}\right)^2 + 2i\pdv{f}{x'}\pdv{\varphi}{x'} + i \pdv[2]{f}{{x'}}\varphi + \pdv[2]{\varphi}{{x'}}\right]e^{if}\\
	\pdv{t} e^{if} \varphi &= \left(i\pdv{f}{t}\varphi + \pdv{\varphi}{t}\right) e^{if}
\end{alignat}
\end{subequations}
where we replaced $t'$ with $t$, as time is universal in Galilean transformations.

By substituting the last formulae (\ref{eq:derivatives}) into~\cref{eq:SchroedingerEGT} we find:
\begin{equation}\label{eq:SchroedingerPhase}
	-\frac{\hbar^2}{2m}\left(\varphi'' - {f'}^2 \varphi + i {f''}^2 \varphi + 2if'\varphi' \right) = i\hbar\left[\left(\dot{\varphi} + i\dot{f}\varphi\right) + \dot{\xi}\left(i{f'}\varphi + {\varphi'}\right)\right]
\end{equation}
where we introduced a new notation for space derivative $f' = \pdv{f}{x'}$ and, as usual, time derivative $\dot{f} =\pdv{f}{t}$.
Firstly, in order to dispose of $\varphi'$, we equate its coefficients on both sides of~\cref{eq:SchroedingerPhase} which gives:
\begin{equation}\label{eq:equating}
	\pdv{f}{x'}\left(x', t\right) = - \frac{m}{\hbar} \dot{\xi} ,
\end{equation}
and by integrating,
\begin{equation}
	f\left(x', t\right) = \frac{m}{\hbar}\left(-\dot{\xi}x'+g\left(t\right)\right)
\end{equation}
where $g$ is an arbitrary function still to be determined.
Secondly, we deduce from~\cref{eq:equating} that the third term of~\cref{eq:SchroedingerPhase} is null:
\begin{equation*}
	\pdv[2]{f}{{x'}} = 0
\end{equation*}
as $\dot{\xi}$ is only a function of $t$.

Now, we can determine the function $g$ by eliminating all extra terms in~\cref{eq:SchroedingerPhase} that are in the form of an explicit time dependence, \ie{} $h(t) \varphi$. Thus $g$ must satisfy:
\begin{equation*}
	\dot{g} = \frac{1}{2} \dot{\xi}^2 \rightarrow g\left(t\right) = \frac{1}{2} \int_{t_0}^t \dot{\xi}^2 \dd{t} .
\end{equation*}
Finally, we come to the new Schrödinger equation, totally equivalent to the previous one of~\cref{eq:Schroedinger}, in an accelerated frame:
\begin{equation}\label{eq:SchroedingerAcc}
\begin{cases}
	\displaystyle
	-\frac{\hbar^2}{2m}\pdv[2]{\varphi}{{x'}} - m\ddot{\xi}x'\varphi = i\hbar\pdv{\varphi}{t} &\\
	\displaystyle
	\psi\left(x',t\right) = e^{i f\left(x',t\right)}\varphi\left(x',t\right), &
	\displaystyle
	f\left(x',t\right) = \frac{m}{\hbar}\left(-\dot{\xi}x' + \frac{1}{2} \int_{t_0}^{t}\dot{\xi}^2 \dd{t}\right) .
\end{cases}
\end{equation}

As we can see, a uniform gravitational field $V_G = - \ddot{\xi} x'$ appears due to the acceleration, which is exactly what one expects from the strong equivalence principle.
However, the mass appears in the phase factor $f$ as well, which could eventually cause mass-dependent diffraction effects.
For this reason, non-relativistic \acl{QM} do not fulfill the strong equivalence principle.

In \ac{QM}, the phase factor is usually related to (proper) time and is actually a measure of it.
For instance, the coefficient of a stationary state for a system with a time-independent Hamiltonian $H$ is $e^{\minus\frac{i}{\hbar}Ht}$, whereas the Feynman path integral form is $e^{\frac{i}{\hbar}\int Ldt}$.
It is possible to prove that the phase factor $f$ of~\cref{eq:phase} is also related to time.

If we consider a particle in a \emph{relativistic} context, one can boost from the initial frame of reference to an other moving with constant velocity $v$ through the Lorentz transformation:
\begin{equation}
	\dd{x'} = \gamma\left(\dd{x} - v\dd{t}\right), \quad \dd{t'} = \gamma\left(\dd{t} - \frac{v}{c^2}\dd{x}\right)
\end{equation}
where $\gamma = \frac{1}{\sqrt{1-v^2/c^2}}$ is the usual Lorentz factor and $c$ is the speed of light.
In the non-relativistic limit $v \ll c$, at the second order approximation, one comes to:
\begin{multline}\label{eq:timeTerm}
	\dd{t'} = \gamma \dd{t} - \frac{v}{c^2}\gamma \dd{x} = \gamma \dd{t} - \frac{v}{c^2}\left(\dd{x'} + \gamma v \dd{t}\right) = \gamma\left(1 - \frac{v^2}{c^2}\right)\dd{t} - \frac{v}{c^2}\dd{x'} \\
	\xrightarrow{v \ll c} \dd{t} - \frac{1}{2}\frac{v^2}{c^2} \dd{t} - \frac{v}{c^2} \dd{x'} = \dd{t} + \var{\tau}.
\end{multline}
The $\var{\tau}$ extra time term of (\ref{eq:timeTerm}) is the differential form of the one in the phase factor $f$ in (\ref{eq:SchroedingerAcc}):
\begin{align}
	\label{eq:extraTimeTerm}
	\var{\tau} =  - \frac{v}{c^2} \dd{x'} - \frac{1}{2}\frac{v^2}{c^2} \dd{t} \stackrel{v \mapsto \minus\dot{\xi}}{=} \frac{1}{c^2}\left(\dot{\xi}\dd{x'} - \frac{1}{2}\dot{\xi}^2 \dd{t}\right) \ \ & \\
	\label{eq:phaseExtended}
	e^{i f\left(x',t\right)} = \exp{-i\frac{m}{\hbar}\left(\dot{\xi}x' - \frac{1}{2}\int_{t_0}^t\dot{\xi}^2 \dd{t}\right)} & .
\end{align}

A proper time term is somehow built into the non-relativistic accelerated Schrödinger equation.
The phase factor $f$ takes into account a residue of proper time  independent of $c$ in the form of the twin-paradox, even though proper time is not recognised in classical physics to any extent, as time $t$ is universal.
This key discrepancy indicates us that, perhaps, proper time should be elevated to a particles internal property.

%***********************************************************	
\section{Proper time as a classical dynamical variable}
%***********************************************************
\label{sec:propertime}

In the last section~\ref{sec:weakandstrong} we saw that the strong equivalence principle in a non-relativistic quantum context leaves a residue of proper time in the phase factor of the wave function.
This incompatibility gives us a physical hint: proper time, \ie{} the time measured with a clock by an observer situated on the particle, should take up full citizenship in \acl{CQM}.
In this section we will examine whether it is possible to interpret proper time as a dynamical variable of a particle, viz. an internal degree of freedom.
Most contents here developed are in the rationale of \cite{Greenberger1970-I}.

First of all, we discuss the physical significance of a \keyword{system internal degrees of freedom}.
On the one hand, they come into play with a mathematical role in the formalism used to describe \emph{all possible configurations} of the system.
The number of independent variables is the total number of degrees of freedom minus the number of constraints applied to the system.
Let us, for example, consider a point mass forced to move on a rod.
We can consider either one position variable, say the distance $x$ from an incision on the rod, or all three $x,y,z$ coordinates with respect to an external point furnished with two holonomic constraints.
In the latter method, the extra complexity involved in the problem set-up rewards us by returning the description of the constraining forces.
Besides their mathematical meaning, the other two coordinates $y,z$ are physical if and only if we are able to remove such constraints (for this instance the rod) and, as a result, the particle is ``free'' to move in all three dimensions.
In the case of observable movements of the particle the additional coordinates acquire physical significance, otherwise they are only a mathematical fiction.

On the other hand, the independent degrees of freedom explicitly provide all necessary information to describe the initial configuration of the system and to predict any subsequent change.
The future behavior must be uniquely determined from its initial state and all previous history has to be completely irrelevant.
Therefore, the mathematical ability to set arbitrary the initial conditions attains any physical significance if we are actually able to interfere with the system, or at least in principle, and its evolution is independent of any procedure required to set it up initially.

In classical non-relativistic physics, proper time has no meaning, as time is universal and the clock on any particle ticks at the same rate.
Only in special relativity, clocks on moving particles run slow and, if they are accelerated, it is not possible to write a single valued function $\tau = \tau(x,v,t)$ of the coordinates in order eliminate proper time.
Supposing that the particle's clock starts synchronized with the laboratory one, they will disagree in a generic motion, even after the particle's clock has come to rest.
However, the laboratory time derivative of proper time $\dot{\tau} = \dv{\tau}{t} = \sqrt{1-\frac{v^2}{c^2}}$ is a function of the coordinates, as the velocity is, and thus the clock rate is uniquely determined at every instant.
That is why proper time can not be an independent degree of freedom in special relativity.

Thanks to the equivalence principle, which leads to the red-shift phenomenon and general relativity, the time rate measured in a position in space depends on the gravitational potential:
\begin{equation*}
	\dd{\tau} = \sqrt{g_{\mu\nu} \dd{x}^\mu \dd{x}^\nu} .
\end{equation*}
Therefore, the rate $\dot{\tau}$ of a particle proper time is merely a measure of the distribution of masses in the universe, which generates the gravitational potential in which the particle is immersed.
In principle, by properly moving the surrounding masses, it will be possible to effect the value of both $\tau$ and $\dv{\tau}{t}$ relative to local coordinate time; than, the temporal evolution of the system will be independent of the past history prior to the initial set-up.

We conclude that, only in a general relativity context it is possible to consider the proper time of a particle as an internal degree of freedom.

%%%%%%%%%%%%%%%%%%%%%%%%%%%%%%%%%%%%%%%%%%%%%%%%%%%%%%%%%%%%
\chapter{The Mass as a Quantum Observable}
%%%%%%%%%%%%%%%%%%%%%%%%%%%%%%%%%%%%%%%%%%%%%%%%%%%%%%%%%%%%
\label{cap:mass}

In non-relativistic classical and \acl{QM}, mass appears as an external and immutable parameter.
Apart from studies on, \eg{}, rockets and motor vehicles, where there is actually a \emph{redistribution} of mass, no theory consider the possibility of (proper) mass changes in the very sense of the word.

In the following chapter we will approach a paradox in \acl{QM}, which arises from the Bargmann theorem and mass superselection rule.
As a consequence of the theorem, it is not possible to have a superposition of different masses for the same particle wavepacket and to keep the Galilean group as the symmetry group of the physical system.
We know that the Galilean group is not the most appropriate symmetry for many reasons; in spite of that, it is not possible that the limit case of low velocities imposes a superselection rule which is not required in the larger symmetry of the Poincaré group.
In order to solve the issue, in the next chapter we will assume that mass is a dynamical variable conjugated to proper time, exactly like position and momentum are.

The following subjects are treated in the articles
\cite{Greenberger1974b,Greenberger2001a,Greenberger2001b}, and in the paper \cite{Greenberger2010a}.

%***********************************************************
\section{The Bargmann theorem and the mass superselection rule in non relativistic \acl{QM}}
%***********************************************************
\label{sec:Bargmann}

In his article \cite{Bargmann1954}, \Bargmann{} evaluates the consequences of a Galilean boost applied to a wave packet made up of states with different mass.
Let us consider a free particle in superposition of the two different masses $m_1$ and $m_2$:
\begin{equation}\label{eq:superpositionMasses}
	\Psi_S(x,t) = \psi_{m_1}(x,t) + \psi_{m_2}(x,t)
\end{equation}
which undergoes the following Galilean transformations from the initial inertial frame $S\left(x,t\right)$:
\begin{center}
\begin{tabular*}{\textwidth}{@{}rl@{\extracolsep{\fill}}l@{}}
	$T_1\colon S\; \mapsto S_1$	& Space translation by $l$:	& $x_1 = x - l$ \\
	$T_2\colon S_1 \mapsto S_2$	& Boost by velocity $v$:	& $x_2 = x_1 - vt$ \\
	$T_3\colon S_2 \mapsto S_3$	& Space translation by $-l$:	& $x_3 = x_2 + l$  \\
	$T_4\colon S_3 \mapsto S_4$	& Boost by velocity $-v$:	& $x_4 = x_3 + vt = x$ \\
\end{tabular*}
\end{center}
with no time translations $t = t_1 = t_2 = t_3 = t_4$ and $l,v\neq 0$.

Following all four changes of reference one comes back to the first one, as their composition is the identity.
For this reasons, one expects the wave function $\Psi$ in the fourth frame being the same of the initial one.
By recalling the transformation of (\ref{eq:EGT}) with $\xi\left(t\right) = \pm vt$, $t_0 = 0$ and for a generic particle of mass $m$ we obtain:
\begin{equation}
	\phi_\pm'\left(x',t\right) = e^{\minus\frac{i}{\hbar}m\left(\pm vx'-\frac{1}{2}v^2t\right)} \phi(x,t)
\end{equation}
that is the formula for the \acl{EGT} in~\cref{eq:SchroedingerEGT,eq:phaseExtended} of~\cref{sec:weakandstrong} with no acceleration and $\varphi = \phi$, as they are solution of the same Schrödinger equation.
Thus:
\begin{align}
	\phi_1(x_1,t) &:= T_1\phi(x_1,t) = \phi(x,t) \\
	\phi_2(x_2,t) &:= T_2\phi_1(x_2,t) = e^{\minus\frac{i}{\hbar}m\left(-vx_2-\frac{1}{2}v^2t\right)} \phi_1(x_1,t) \\
	\phi_3(x_3,t) &:= T_3\phi_2(x_3,t) = \phi_2(x_2,t) \\
	\phi_4(x_4,t) &:= T_4\phi_3(x_4,t) = e^{\minus\frac{i}{\hbar}m\left(+vx_4-\frac{1}{2}v^2t\right)} \phi_3(x_3,t) .
\intertext{Combining them all up, the wave function in $S_4$ reads:}
	\phi_4(x_4,t) &= e^{\minus\frac{i}{\hbar}mvl}\phi(x_4,t) \equiv e^{\minus\frac{i}{\hbar}mvl}\phi(x,t) .
\end{align}
As we can see, a phase function of the mass $m$ is introduced; still, it is unmeasurable and therefore non physical.

Things are different for our wavepacket in a \emph{coherent} superposition of two masses (\ref{eq:superpositionMasses}). We have in the last frame of reference:
\begin{equation}\label{eq:phaseShift}
\begin{split}
	\Psi_{S_4}(x_4,t) &= e^{\minus\frac{i}{\hbar}m_1 vl}\psi_{m_1}(x_4,t) + e^{\minus\frac{i}{\hbar}m_2 vl}\psi_{m_2}(x_4,t) \\
	&= e^{\minus\frac{i}{\hbar} m_1 vl}\left[\psi_{m_1}(x_4,t) + e^{\minus\frac{i}{\hbar} \DeltaVar{m} vl}\psi_{m_2}(x_4,t)\right]
\end{split}
\end{equation}
where $\Delta m = m_2 - m_1$.
The composition of transformations leaves a measurable and physical phase shift $e^{\minus\frac{i}{\hbar}\DeltaVar{m}vl}$ in the last frame of reference between the two terms in superposition due to the different masses.
Nevertheless, non-relativistically the first and the fourth frame of reference are exactly the same and hence, the phase shift should be zero.
Equivalently, the state after the four changes of reference frame contains a physical and measurable \emph{memory} of the transformation $e^{\minus\frac{i}{\hbar} \DeltaVar{m} vl}$ and, for this reason, the Galilean group is not the symmetry group for the system.

These consideration are summarized in the following theorem:
\begin{theorem*}[Bargmann]\index{theorem!Bargmann}
	For a quantum system, being invariant under the Galilean symmetry group is incompatible with pure states arising from combinations of different masses.
\end{theorem*}
For further readings and for a more straightforward proof by means of representation theory see \cite[p.~601]{Moretti2013}.
In order to eliminate the extra phase shift in~\cref{eq:phaseShift}, \Bargmann{} proposes to set $\Delta m = 0$ and introduces consequently the following superselection rule:
\begin{massdef*}\index{mass superselection rule}
	In \acl{CQM}, pure states correspond to vectors in Hilbert spaces with definite mass.
\end{massdef*}

When we consider an isolated system which is asked to be invariant under the Galilean group, \ie{} in non-relativistic mechanics, besides energy and momentum conservations we require separately the mass to conserve \emph{and} to be invariant under change of reference frame.
However, with the introduction of mass-energy equivalence in relativity, not only such invariance is lost, but it is quite usual to superimpose states of different masses (or equivalently, different energies).

%***********************************************************	
\section{Paradox from Bargmann theorem}
%***********************************************************
\label{sec:BargmannParadox}

The Bargmann theorem becomes unfounded as soon as we take into account special relativity.
With the introduction of the mass-energy equivalence first by A.~Einstein \cite{Einstein1905}, it is no longer possible to treat differently a particle mass and its energy.
Since mass is just another form of energy, one can certainly superimpose states of different masses, as one actually does in many decay processes.
\Greenberger{} claims that a \emph{paradox}\index{paradox!Bargmann} therefore arises, because a grater symmetry, namely Lorentz invariance, produces in the limit of low velocities, \ie{} in the less symmetrical Galilean invariance, an additional (super)selection rule.
Commonly, it is the other way round: a larger symmetry generates an extra rule over the quantum pure states, initially not observable from the original one.

If the thesis of the Bargmann theorem is inconsistent, the cause are to be sought in its hypotheses: Galilean invariance and Schrödinger equation.
In the theory described in \cref{cap:theory}, the Galilean group will be extended to comprehend proper time and therefore preserve the validity of the Schrödinger equation.

A clue on the paradox physical significance can be sought by observing that the extra phase shift in~\cref{eq:phaseShift} is nothing but the difference in proper time elapsed between the two systems.
Let us consider the same changes of reference as in~\cref{sec:Bargmann} relativistically, by means of the Lorentz transformation:
\begin{center}
\begin{tabular*}{\textwidth}{@{}rl@{\extracolsep{\fill}}ll@{}}
	$\overline{T_1}\colon S\; \mapsto S_1$	& Space translation by $l$:	& $x_1 = x - l$ & $t_1 = t$ \\
	$\overline{T_2}\colon S_1 \mapsto S_2$	& Boost by velocity $v$:	& $x_2 = \gamma\left(x_1 - vt_1\right)$ & $t_2 = \gamma\left(t_1 - \frac{v}{c^2}x_1\right)$ \\
 	$\overline{T_3}\colon S_2 \mapsto S_3$	& Space translation by $-l$:\footnotemark	& $x_3 = x_2 + \frac{l}{\gamma}$ & $t_3 = t_2$ \\
 	$\overline{T_4}\colon S_3 \mapsto S_4$	& Boost by velocity $-v$:	& $x_4 = \gamma\left(x_3 + vt_3\right)$ & $t_4 = \gamma\left(t_3 + \frac{v}{c^2}x_3\right)$ \\
\end{tabular*}
\end{center}
\footnotetext{NB: in order to go back at the original point, the spatial translation is by $-\frac{l}{\gamma}$ because of Lorentz space contraction.}
where $\gamma = \frac{1}{\sqrt{1-v^2/c^2}}$ is the usual Lorentz factor.
By composing them together we come to:
\begin{gather}
	\label{eq:finalCoordsEGT}
	x_4 = x, \quad t_4 = t + l\frac{v}{c^2} \\
	\Delta\tau = t_4 - t = l\frac{v}{c^2}
\end{gather}
which leads naturally to the phase shift of $e^{\minus\frac{i}{\hbar}\DeltaVar{m} c^2 \Delta\tau} = e^{\minus\frac{i}{\hbar}\DeltaVar{m} vl} $.

The unexpected result is that the difference in proper times between the reference frames produces a residue in non relativistic physics totally independent of $c$.
In a special relativity context it is a physically meaningful effect, whereas in classical mechanics it has no adequate interpretation.
Just settling things once and for all by prohibiting superposition states of different masses without any physical reasoning may mislead to incomplete theories.
Thus, we shall not follow Bargmann's solution.

The previous analysis is in many details very similar to the results of~\cref{sec:weakandstrong} and, for this reason, it further supports the argument of proper time as dynamical variable.

%***********************************************************	
\section{Mass as a canonical variable}
%***********************************************************
\label{sec:massvariable}

The role of mass in \acl{QM} has been inherited directly from the classical Newtonian theory: it is an external constant parameter with no dynamical behaviour.
Even in relativistic mechanics, where energy and mass undergo an equivalence relation, there is no way for the mass parameter to follow dynamically the energy variations due to interactions between different systems.
One has to adjust it by hand.

As we saw in \cref{sec:weakandstrong}, the geometrical interpretation of the equivalence principle is that no additional parameter is required in order to describe the trajectory of a particle in a gravitational field.
However, the substitution of (\ref{eq:masslessQuantities}) becomes pointless as soon as we consider a \emph{non-gravitational} interaction.
E.~g., for the electromagnetic field the energy potential is no longer linear in the mass:
\begin{align*}
	V_G \left(r\right) = G\frac{M}{r} & \longleftrightarrow V_C \left(r\right) = \frac{Q}{r} \\
	H = \frac{1}{2}mv^2 + mV_G & \longleftrightarrow H =\frac{1}{2}mv^2 + qV_C
\end{align*}
and one is not able to group the mass out.
The mass parameter $m$ is required and meaningful only for approaching systems which interact non-gravitationally.
A free isolated particle has no mass as long as it does not interact.

\Greenberger{} in his paper \cite{Greenberger2010a} suggests an example:
\begin{quote}
	If a particle decays into two, of known rest masses $m_1$ and $m_2$, one does not know the masses of either one until it interacts with something.
	For example, if one of them passes through a slit, it acquires a de Broglie wavelength, which is mass-dependent.
	It is only then that one knows which type of particle passed through the slit.
\end{quote}
In the case of a gravitational interaction instead, we would not be able to measure their masses because of the equivalence principle.

Assuming that a particle owns a definite mass at all times even without having measured it, or else, after having measured it, taking for granted that it has always had this mass, it is actually against any quantum principle.
In his paper \cite{Greenberger2010a}, \Greenberger{} wonders the reason of why should not a particle have a definite value for its spin until it has been measured, but have always a well-defined mass.
This is actually EPR thinking and we could, in fact, formulate a Bell theorem for the case.

If a particle acquires its mass only by means of an interaction, it must follow the same rules as for all quantum observables.
A self-adjoint linear operator is needed to carry out a \emph{measure} on the particle.
The theory formalism and its consequences will be developed in the next chapter.

%% Secondo aggiornamento per arXive
The same conclusions of \Greenberger{} have been drawn more recently and independently in \cite{Giulini1996,Annigoni2013}, where the authors employ the far more rigorous formalism of group representations to inspect the physical significance of the mass super selection rule.
In particular, D.~Giulini states that requiring a superselection rule over mass while interpreting it as parameter do not stand to reason.
Therefore he extends the usual Galilean symmetry to include the mass as a variable, that is what we will do in the next chapter, and only then he discusses the consequences of the selection rule.
The main difference between these articles and \Greenberger{}'s theory is that they do not make any assumption about the physical meaning of the variable conjugated to the mass.

%%%%%%%%%%%%%%%%%%%%%%%%%%%%%%%%%%%%%%%%%%%%%%%%%%%%%%%%%%%%
\chapter{A Theory of Particles with Variable Mass}
\chaptermark{Theory of Particles with Variable Mass}
%%%%%%%%%%%%%%%%%%%%%%%%%%%%%%%%%%%%%%%%%%%%%%%%%%%%%%%%%%%%
\label{cap:theory}

What we have developed in the first chapters suggests and enables us to introduce a theory with new degrees of freedom.
A non-relativistic boost applied to the Schrödinger equation points to the lack of a proper time definition in \acl{CQM}; moreover, the paradox that arises from Bargmann's theorem invite us to wonder if being in a superposition of different masses should lie in particles' power even at low velocities.

In the following chapter we will address the formalism and consequences of a classical and \acl{QM} developed by \Greenberger{} in which mass and proper time are assumed as generalized canonically conjugated coordinates (conjugated observables after quantization).

The theory was first divulged in \cite{Greenberger1970-I,Greenberger1970-II} and some consequences are discussed in \cite{Greenberger1974a,Greenberger1974b}.
A few arguments are recalled eventually in \cite{Greenberger2010a}.

%***********************************************************
\section{Classical theory}
%***********************************************************
\label{sec:theory-classical}

In order to set up a classical theory of particles with variable mass, we shall consider a pointlike particle described by the coordinates position $q_i$, proper time $\tau$ and their velocities $\dot{q_i},\dot{\tau}$.
From now on, proper time acquires full citizenship as a generalized coordinate, exactly like $q_i$, and its momentum conjugate is the particle mass.\footnote{The conjugate momentum of proper time is actually $mc^2$, which has the units of energy.}
Thus we have:
\begin{gather}
	L = L\left(q_i,\dot{q_i},\tau,\dot{\tau},t\right) \\
	\label{eq:ConjugatedMomenta}
	p_i = \pdv{L}{\dot{q_i}}, \quad mc^2 = \pdv{L}{\dot{\tau}}
\end{gather}
where $L$ is the Lagrangian and $c$ the speed of light.
Henceforth, we will further assume that all the coordinates $q_i$ are independent of the mass and the momenta $p_i$ are linear to it, namely
\begin{equation}\label{eq:hypMomentaLinear}
	p = m f \left(q_i,\dot{q_i}\right) .
\end{equation}

We can then write down a Hamiltonian for our particle:
\begin{equation}\label{eq:Hamiltonian}
	H\left(q_i,p_i,\tau,m,t\right) = p_i\dot{q_i} + c^2m\dot{\tau} - L
\end{equation}
equipped with the Hamilton's equations
\begin{equation}\label{eq:HamiltonianEq}
\begin{split}
	\dot{q_i} = \pdv{H}{p_i}, \quad & \dot{p_i} = - \pdv{H}{q_i} \\
	\dot{\tau} = \frac{1}{c^2}\pdv{H}{m}, \quad & \dot{m} = - \frac{1}{c^2}\pdv{H}{\tau} .
\end{split}
\end{equation}
The coordinates are indeed canonically conjugated:
\begin{gather}\label{eq:classicalPoisson}
	\pb{x}{p} = 1 \qcomma \pb{\tau}{mc^2} = 1 \\
\intertext{where}
	\pb{f}{g} = \left(\pdv{f}{x_i}\pdv{g}{p_i} - \pdv{f}{p_i}\pdv{g}{x_i}\right) + \frac{1}{c^2}\left(\pdv{f}{\tau}\pdv{g}{m} - \pdv{f}{m}\pdv{g}{\tau}\right) .
\end{gather}
are the Poisson's brackets.

As a consequence of our hypotheses in \cref{eq:hypMomentaLinear}, the Hamiltonian is a homogeneous function of the first order in $p_i$ and $m$, \ie{} $H(q_i,\alpha p_i,\tau,\alpha m,t) = \alpha H(q_i,p_i,\tau,m,t)$.
By virtue of the Euler's theorem for a linear homogeneous function, one comes to:
\begin{equation}
	H = p_i \pdv{H}{p_i} + m \pdv{H}{m}
\end{equation}
but, from \cref{eq:HamiltonianEq} we yield:
\begin{equation}\label{eq:HamiltonianHomogeneus}
	H = p_i\dot{q_i} + c^2 m\tau .
\end{equation}
Last \cref{eq:HamiltonianHomogeneus} tells us that the energy of the particles is linear to the mass.
According to what we have already saw in \cref{sec:weakandstrong} this is consistent with the weak equivalence principle; we did not lost any feature from standard classical physics.
By comparing \cref{eq:HamiltonianHomogeneus} to \cref{eq:Hamiltonian} we deduce that the weak equivalence is satisfied if and only if the Lagrangian is $L = 0$.

\subsection{Relativistic free particle}
Let us consider the usual relativistic Hamiltonian for a free particle with $\left(x,\tau\right)$ coordinates:
\begin{equation}\label{eq:HamiltonianRelativisticVariableMass}
	H = \sqrt{p^2c^2 + m^2c^4} \stackrel{c = 1}{=} \sqrt{p^2 + m^2}.
\end{equation}
This simple example illustrates how naturally the new theory extends the conventional one.
Some great results are achieved by evaluating the Hamilton's equations of motion (\ref{eq:HamiltonianEq}) for the case:
\begin{equation}
\begin{split}\label{eq:HamiltonianEqFreeRelPar}
	\dot{q} = \frac{pc^2}{\sqrt{p^2c^2 + m^2c^4}}, & \quad \dot{p} = 0 \\
	\dot{\tau} = \frac{mc^2}{\sqrt{p^2c^2 + m^2c^4}}, & \quad \dot{m} = 0 . \\
\end{split}
\end{equation}
By accordingly inverting the first equation of (\ref{eq:HamiltonianEqFreeRelPar}) to solve in terms of $p$ and reminding that $v = \dot{q}$ one comes to:
\begin{equation}\label{eq:HamiltonianEqSub}
	p = \frac{mv}{\sqrt{1 - v^2/c^2}}, \quad\dot{\tau} = \sqrt{1 - v^2/c^2} .
\end{equation}
The last two equation (\ref{eq:HamiltonianEqSub}) were once part of space-time geometry, in which mechanics described evolutions.
They have now become part of the dynamics of the system, as they are equations of motion.

As in conventional relativity, in our theory the mass $m$ is not the invariant mass $m_0$; it is actually the \emph{energy} of the particle in its rest frame, that is the frame in which the momentum is zero.
Thus, in general it must include the binding energies, which arise from gravitational and non-gravitational potentials.

\subsection{Classical particle in a gravitational potential}\label{subsec:classicalGrav}
The previous result, namely that $L = 0$ in order to satisfy the weak equivalence principle, serves us to construct a Lagrangian for the limit case of a non-relativistic particle immersed in an external \emph{gravitational} potential\footnote{In this chapter the gravitational potential will be denoted with $\varphi$ instead of $V_G$ as in \cref{sec:weakandstrong} for notation simplicity.} of the form $m\varphi(x_i,\tau,t)$.
For an arbitrary function $\sigma = \sigma\left(x_i,\tau,\dot{x}_i,\dot{\tau}\right)$, we have at the second order approximation:
\begin{gather}
	L = \sigma c^2\left(\dot{\tau} - \sqrt{g_{\mu\nu}\dot{x}^\mu\dot{x}^\nu}\right) \stackrel{v \ll c}{\simeq} \sigma c^2\left[\dot{\tau} - \left(1 - \dfrac{1}{2}\frac{\vb{v}^2}{c^2} + \frac{\varphi}{c^2}\right)\right] \equiv \sigma c^2 \Xi \\
	\Xi\left(x_i,\tau,\dot{x}_i,\dot{\tau}\right) = \dot{\tau} - 1 + \dfrac{1}{2}\frac{\vb{v}^2}{c^2} - \frac{\varphi}{c^2}
\end{gather}
in which $\vb{v}$ is the particle speed. From \cref{eq:ConjugatedMomenta} for the momenta we obtain:
\begin{equation}
	p_i = \pdv{L}{\dot{x}_i} = \sigma\dot{x}_i + \Xi \pdv{\sigma}{\dot{x}_i} = \sigma\dot{x}_i, \quad mc^2 = \pdv{L}{\dot{\tau}} = c^2\left(\sigma + \Xi\pdv{\sigma}{\dot{\tau}}\right) = \sigma c^2
\end{equation}
where we assumed $\Xi = 0$ to satisfy $L = 0$ and let $m = \sigma$ not necessarily constant.
The Euler–Lagrange equations are:
\begin{align}
	\label{eq:ELeqParticleMomentum}
	\dot{p}_i &= \pdv{L}{x_i} = - \pdv{\sigma}{x_i} c^2 \Xi - \sigma\pdv{\varphi}{x_i} = - m\pdv{\varphi}{x_i} \\
	\label{eq:ELeqParticleMass}
	c^2\dot{m} &= \pdv{L}{\tau} = \pdv{\sigma}{\tau} c^2 \Xi - \sigma\pdv{\varphi}{\tau} = - m\pdv{\varphi}{\tau}
\end{align}
which indicates that any variation of the mass $m$ relies on the $\tau$-dependence of the gravitational potential $\varphi$.
We can now write the Hamiltonian for the particle from \cref{eq:Hamiltonian}:
\begin{equation}\label{eq:HamiltonianVariableMass}
	H = mc^2 + \frac{p^2_i}{2m} + m\varphi
\end{equation}
and the Hamilton's equations are:
\begin{equation}\label{eq:HamiltonEqParticle}
\begin{split}
	\dot{x}_i = \pdv{H}{p_i} = \frac{p_i}{m}, \quad & \dot{p_i} = - \pdv{H}{x_i} = - m \pdv{\varphi}{x_i} \\
	\dot{\tau} = \frac{1}{c^2}\pdv{H}{m} = 1 - \dfrac{1}{2}\frac{\vb{v}^2}{c^2} + \frac{\varphi}{c^2}, \quad & \dot{m}c^2 = - \pdv{H}{\tau} = - m \pdv{\varphi}{\tau} .
\end{split}
\end{equation}
If we combine \cref{eq:HamiltonEqParticle} with \cref{eq:ELeqParticleMomentum} we have:
\begin{equation}
	\dot{p}_i = m\ddot{x}_i + \dot{x}_i\dot{m} = - m\pdv{\varphi}{x_i} \longrightarrow\ddot{x}_i = \frac{1}{c^2}\pdv{\varphi}{\tau}\dot{x}_i - \pdv{\varphi}{x_i}
\end{equation}
where the $\dot{x}_i$ term is a $1/c^2$ correction to the motion due to the changing mass.
Again, we come to an equation of motion totally independent of the mass $m$, and thus, we have not lost any property of the conventional mechanics.

We remark that in the article \cite{Greenberger1970-I}, a relativistic gauge field theory has been developed.
Even though it will not be described in this dissertation, we will assume that our results are the non-relativistic limit case of the theory.
Hence, not all gravitational potential $\varphi$ are allowed, but they have to satisfy the following gauge condition:
\begin{equation}\label{eq:gaugeCondition}
	\pdv{\varphi}{t} + \frac{g}{c^2}\varphi\pdv{\varphi}{\tau} = 0
\end{equation}
where $g$ is the coupling constant.
A more detailed description is available in \cref{app:gaugeCondition}.

\subsection{Classical decay theory}\label{subsec:decay}
One of the most relevant feature of the theory is that it enables the particle decay phenomena to be described classically.
If we take into account the Hamiltonian of \cref{eq:HamiltonianVariableMass} for a particle which remains at rest, \ie{} for which no position-dependent forces act, we have:
\begin{gather}
	H = mc^2 + m\varphi \\
	\label{eq:decayProperTimeEq}
	\dot{\tau} = 1 + \frac{\varphi}{c^2}
\end{gather}
for a potential $\varphi = \varphi(\tau,t)$ still to be determined.
By accordingly solving the gauge condition of \cref{eq:gaugeCondition} (see \cref{app:gaugeCondition}) we have
\begin{equation}
	\dot{\varphi} = \pdv{\varphi}{\tau} \tag{\ref{eq:gaugeConditionResult}}
\end{equation}
which leads, from the mass $m$ Hamilton's equation of \cref{eq:HamiltonEqParticle}, to:
\begin{equation}\label{eq:decayGaugeStep}
	\dot{m}c^2 = -m\pdv{\varphi}{\tau} = - m\dot{\varphi} \longrightarrow \frac{\dot{m}}{m} = - \frac{1}{c^2}\dot{\varphi}
\end{equation}
so that by integrating
\begin{equation}\label{eq:decayMassSolution}
	m\left(t\right) = m_0 \exp(-\frac{\varphi(t) - \varphi(t_0)}{c^2}) .
\end{equation}
The mass $m$ solution inherits the time dependence of the potential $\varphi$ as already remarked next to \cref{eq:ELeqParticleMass}.
For instance, we can solve for $\varphi$ starting from a negative exponential decay:
\begin{equation}
	m\left(t\right) = m_0 + \mu e^{\minus\gamma t}
\end{equation}
where the particle initial mass ($t_0 = 0$) is $m_0 + \mu$ and $m = m_0$ at $t \to + \infty$ .
Then we have:
\begin{equation}
	\frac{\dot{m}}{m}(t) = - \frac{\gamma\mu e^{\minus\gamma t}}{m_0 + \mu e^{\minus\gamma t}} \stackrel{\mu\ll m_0}{\simeq} - \frac{\gamma\mu e^{\minus\gamma t}}{m_0} = - \frac{1}{c^2}\dot{\varphi}
\end{equation}
thus,
\begin{equation}\label{eq:gravPotTime}
	\varphi\left(t\right) \simeq - \frac{\mu c^2}{m_0} e^{\minus\gamma t} .
\end{equation}
where we assumed $\varphi\left(0\right) = 0$ in order to normalize the potential to zero at $t\to +\infty$ and have $\dot{\tau} = 1$ according to \cref{eq:decayProperTimeEq}.

The result of \cref{eq:gravPotTime} manifests an innovative feature of the theory.
If we consider \cref{eq:HamiltonEqParticle} for a decaying particle at rest ($\vb{v} = 0$), we have
\begin{equation}
	\dot{\tau} = 1 + \frac{\varphi}{c^2} \simeq 1 - \frac{\mu}{m_0} e^{\minus\gamma t}
\end{equation}
and by integrating
\begin{equation}
	\tau(t) \simeq t - \int_{0}^{t}\frac{\mu}{m_0} e^{\minus\gamma t}\dd{t} = t - \frac{\mu}{\gamma m_0}\left(1 - e^{\minus\gamma t}\right) \xrightarrow{t \to +\infty} t - \frac{1}{\gamma}\frac{\mu}{m_0} .
\end{equation}
The rate at which proper time changes for a decaying particle disagrees with the laboratory one.
It means that, for a stable particle, proper time evolves differently than for an unstable one, even thought they are both at rest.
In the particular case in which the decaying system irradiates, the frequency of the emitted electromagnetic wave measured in the laboratory frame of reference will differ from that of the particle.
Even for a pointlike particle, ergo elementary and without microscopic substructure, our theory predicts a \keyword{decay red shift} phenomenon.
\todo{Principio equivalenza potenziale dopo accelerazione p.404 Properties}

\subsection{Binding energy as inertia}\label{subsec:bindingEnergy}
According to relativity, if a particle interacts with an external force field and changes its energy, the inertia should vary as well.
In conventional mechanics, however, there is no way for the particle to keep pace with the external interactions and one has to correct it by hand.
Things are different in our theory, where mass is free to change in agreement with the equations of motion. 

If we consider a free particle of mass $m_0$ that suddenly interacts with an external \emph{non-gravitational} potential $V = V(x_i,t)$, a $\tau$-dependent \emph{gravitational} potential $\varphi$ appears, whose role is to change the mass.
As soon as it reaches the correct value consistent with the energy, kinetic \emph{and} binding, the potential $\varphi$ disappears and the particle evolves with the new mass $\mu_o$.

In the case of a potential $V$, one simply adds it to the Hamiltonian of \cref{eq:HamiltonianVariableMass}:
\begin{equation}
	H = mc^2 + \frac{p_i^2}{2m} + m\varphi + V .
\end{equation}
which leads to the following Hamilton's equations of motion:
\begin{equation}\label{eq:HamiltonEquationPhiV}
\begin{split}
	\dot{x}_i = \frac{p_i}{m}, \quad & \dot{p_i} = - m \pdv{\varphi}{x_i} - \pdv{V}{x_i} \\
	\dot{\tau} = 1 - \dfrac{1}{2}\frac{\vb{v}^2}{c^2} + \frac{\varphi}{c^2}, \quad & \dot{m}c^2 = - m\pdv{\varphi}{\tau} .
\end{split}
\end{equation}
The gauge condition for $\varphi$ in \cref{eq:gaugeCondition} has the same result as in the previous paragraph, hence \cref{eq:gaugeConditionResult,eq:decayGaugeStep,eq:decayMassSolution} are still valid.
Furthermore we have
\begin{equation}
	\pdv{\varphi}{x_i} = \frac{1}{2} \frac{\dot{x}_i}{c^2}\pdv{\varphi}{\tau} . \tag{\ref{eq:bindingGaugeResult}}
\end{equation}
If we consider the first two equation of (\ref{eq:HamiltonEquationPhiV}) and \cref{eq:bindingGaugeResult} we can write:
\begin{equation}
	\dot{p_i} = m\ddot{x}_i + \dot{x}_i\dot{m} = - m \pdv{\varphi}{x_i} - \pdv{V}{x_i} = - \frac{1}{2}m\frac{\dot{x}_i}{c^2}\pdv{\varphi}{\tau} - \pdv{V}{x_i} ,
\end{equation}
by virtue of \cref{eq:gaugeConditionResult} and the third of (\ref{eq:HamiltonEquationPhiV}) we come to
\begin{equation}\label{eq:bindingStep1}
	\ddot{x}_i - \frac{1}{2}\frac{\dot{\varphi}}{c^2}\dot{x}_i = - \frac{1}{m}\pdv{V}{x_i} .
\end{equation}
Now, we group the first member of \cref{eq:bindingStep1} and substitute for \cref{eq:decayMassSolution} to attain
\begin{gather}
	e^{\frac{1}{2}\frac{\varphi}{c^2}} \dv{t}\left(e^{\minus\frac{1}{2}\frac{\varphi}{c^2}}\dot{x}_i\right) = - \frac{\exp(\frac{\varphi\left(t\right) - \varphi\left(t_0\right)}{c^2})}{m_0}\pdv{V}{x_i} \\
	m_0 e^{\frac{\varphi\left(t_0\right)}{c^2}} e^{\minus\frac{1}{2}\frac{\varphi}{c^2}} \dv{t}\left(e^{\minus\frac{1}{2}\frac{\varphi}{c^2}}\dot{x}_i\right) = - \pdv{V}{x_i} .
\end{gather}

Then, if we define
\begin{equation}
	\lambda\left(t\right) = \int_{t_0}^t e^{\frac{1}{2}\frac{\varphi}{c^2}} \dd{t} \qcomma \dv{\lambda} = e^{\minus\frac{1}{2}\frac{\varphi}{c^2}} \dv{t}
\end{equation}
we come to a reduced equation of motion
\begin{equation}\label{eq:bindingEquationReduced}
	m_0 e^{\frac{\varphi\left(t_0\right)}{c^2}} \, \dv[2]{x_i}{\lambda} = - \pdv{V}{x_i}
\end{equation}
in terms of the new time coordinate $\lambda$.
As soon as the particle reaches its final mass, we expect the potential $\varphi$ to vanish:
\begin{equation}
	\dot{\lambda} \xrightarrow{t \to +\infty} 1 \qcomma \dd{\lambda} \xrightarrow{t \to +\infty} \dd{t}
\end{equation}
and \cref{eq:bindingEquationReduced} becomes the usual Newton's equation with final mass $\mu_0 = m_0 e^{\frac{\varphi\left(t_0\right)}{c^2}}$.
In a non-relativistic and non-interacting theory, the initial value for the potential $\varphi\left(t_0\right)$ is totally arbitrary.
For consistency, one may set it so that the mass difference is exactly the binding energy due to $V$, but this is not required.
Exclusively in a more advanced theory in which different interacting system are described, the relationship between the generated potential $\varphi$ and the particle could be demonstrated.
%\todo{Preaccelation?}

%***********************************************************	
\section{Consequences of canonical quantization}
%***********************************************************
\label{sec:theory-quantum}

Starting from the classical Hamiltonian description of the last section \ref{sec:theory-classical}, one is able to evaluate the consequences of a quantum theory for particles with variable mass by means of the \keyword{canonical quantization}.
Hence, from the conjugation in \cref{eq:classicalPoisson} we come to
\begin{align}
	\pb{x}{p} = 1 &\longrightarrow \comm{\widehat{x}}{\widehat{p}} = i\hbar \\
	\pb{\tau}{mc^2} = 1 &\longrightarrow \comm{\widehat{\tau}}{\widehat{m}c^2} = i\hbar
\end{align}
where the notation $\widehat{x}$ distinguishes the quantum observables from the classical variables.
As remarked at the beginning of \cref{sec:theory-classical}, by following these assumptions, mass $m$ and proper time $\tau$ are system internal degrees of freedom exactly as position and momentum are.
Therefore, the related linear operators are used to carry out measures on states.

The first result of quantization is a \emph{mass-proper time} uncertainty relation\index{uncertainty relation!mass-proper time}.
As for the Heisenberg's formula, which states for two observables $A, B \in\mathscr{L}\left(\mathcal{H}\right)$ and $\psi\in\mathcal{H}$:
\begin{equation}
	\DeltaVar{A}\DeltaVar{B} \geq \frac{1}{2} \left|\expval{\comm{A}{B}}{\psi}\right|
\end{equation}
where $\Delta A = \sqrt{\expval{A^2} - \expval{A}^2}$, we have at study
\begin{equation}\label{eq:massProperTimeUncertainty}
	c^2 \DeltaVar{m}\DeltaVar{\tau} \geq\frac{\hbar}{2} .
\end{equation}

It is fundamental, not to mistake \cref{eq:massProperTimeUncertainty} for the common mass-energy uncertainty relation:
\begin{equation}\label{eq:energyTimeUncertainty}
	\DeltaVar{E}\DeltaVar{t} \geq \frac{\hbar}{2} .
\end{equation}
In our theory, proper time and mass are physical quantities measured in a particular system of reference: the particle rest frame.
Therefore, the proper time uncertainty $\DeltaVar{\tau}$ expresses the standard deviation of measures from external frames of the reading an imaginary clock situated on the particle.
The $\DeltaVar{t}$ term of \cref{eq:energyTimeUncertainty}, in its place, is the time interval measured in a particular coordinate system.
It is not only a matter of reference frames, the usual uncertainty originates generally from the kinetic or potential energy of the body, rather than its rest energy.
Whereas the uncertainty of \cref{eq:massProperTimeUncertainty} can arise from both an uncertain gravitational potential or an uncertain velocity through the Lorentz factor $\gamma$.

Some examples of systems and experiments for which the mass-proper time inequality is quite reasonable are illustrated in \cite{Greenberger1970-II,Greenberger1974b,Greenberger2010a}.

A free particle in the quantum theory, like in conventional \ac{QM}, is described as a wave packet:
\begin{equation}
	\psi(\vb{x},\tau,t) = \braket{x,\tau}{\psi} = \frac{1}{\left(2\pi\right)^2} \int a(\vb{k},m) e^{i\left(\vb{k}\vdot\vb{x} + mc^2\tau/\hbar - \omega t\right)} \dd[3]{\vb{k}}\dd{m}
\end{equation}
where $a$ is the distribution in momenta and mass.
In position representation we can define the mass linear self-adjoint operator as follows:
\begin{equation}\label{eq:massOperator}
	mc^2 = -i\hbar\pdv{\tau} \longleftrightarrow m = - i \frac{\hbar}{c^2} \pdv{\tau}
\end{equation}
with which we are able to formulate the relativistic Schrödinger equation:
\begin{equation}
	\left(m^2c^4 - \hbar^2c^2\nabla^2\right)\psi = - \hbar^2 \pdv[2]{\psi}{t} .
\end{equation}

In the presence of an external gravitational potential $\varphi$, we substitute non-relativistically the momentum with
\begin{equation}
	\partial_\mu \longrightarrow \left(\pdv{t} + \varphi\pdv{\tau},\nabla\right)
\end{equation}
and have
\begin{equation}\label{eq:theorySchroedingerEq}
	\left(mc^2 + \frac{p^2}{2m} + V\right)\psi = i\hbar\pdv{\psi}{t} -\varphi m\psi
\end{equation}
or equivalently
\begin{equation}\label{eq:theorySchroedingerEqGrav}
\left(m^2c^2 + \frac{1}{2}p^2 + m \varphi m + mV\right)\psi = i\hbar m\pdv{\psi}{t} .
\end{equation}

%***********************************************************
\section{Solution of the paradox from Bargmann theorem}
%***********************************************************
\label{sec:theory-solution}

We are now able to solve the paradox that arises from Bargmann theorem as described in \cref{sec:Bargmann,sec:BargmannParadox}.
In our theory the mass superselection rule becomes pointless in the non-relativistic quantum theory as the same effect evaluated for the relativistic case are now present even at low velocities.
Let us consider a particle wave-function $\psi = \psi(\vb{x},\tau,t)$ and a change of reference frame from the initial inertial one to an other by means of the \acl{EGT} of \cref{eq:EGT}.
Then, at any time $t$ we have from \cref{eq:SchroedingerAcc};
\begin{equation}
	\psi'\left(\vb{x}',\tau,t\right) = e^{\minus i\frac{m}{\hbar}\left(\dot{\vb*{\xi}}\vdot\vb{x}' - \frac{1}{2}\int_{t_0}^t\dot{\vb*{\xi}}^2 \dd{t}\right)} \psi(\vb{x},\tau,t) .
\end{equation}
In the particular case for which there exists a time $T > t_0 = 0$ so that the general transformation $\vb*{\xi}$ satisfies:
\begin{equation}
	\vb*{\xi}(t) = 0 \quad \dot{\vb*{\xi}}(t) = 0 \quad \ddot{\vb*{\xi}}(t) = 0 \qcomma{\forall t \geq T}
\end{equation}
it is possible to demonstrate by substituting $m$ for the operator of \cref{eq:massOperator} that
\begin{multline}\label{eq:paradoxSolution}
	\psi'(\vb{x},\tau,t)
	= e^{i\frac{1}{2}\frac{m}{\hbar}\int_{0}^T\dot{\vb*{\xi}}^2 \dd{t}} \psi(\vb{x},\tau,t)
	= e^{\frac{1}{2}\int_{0}^T\frac{\dot{\vb*{\xi}}^2}{c^2}\dd{t}\pdv{\tau}} \psi(\vb{x},\tau,t) \\
	= \psi\left(\vb{x},\tau + \frac{1}{2}\int_{0}^T\frac{\dot{\vb*{\xi}}^2}{c^2}\dd{t},t\right) \qcomma{\forall t\geq T}
\end{multline}
and therefore the transformation does not reveal any inconsistency.
The additional term for the wavefunction proper time takes into account the physical difference in time rate between an inertial and an accelerated frame of reference.
As we have already seen in \cref{sec:weakandstrong,sec:BargmannParadox}, the term in \cref{eq:SchroedingerAcc}, and as a consequence in \cref{eq:paradoxSolution}, is the integration over time for the Lorenz time contraction in the non-relativistic limit:
\begin{equation}
	\dd{\tau} = \sqrt{1 - \frac{v^2}{c^2}}\dd{t} \stackrel{v \ll c}{\simeq} \left(1 - \frac{1}{2}\frac{v^2}{c^2}\right) \dd{t} .
\end{equation}
For this reason the formalism produces exactly the time difference required by relativity, hence being consistent.

We conclude that the greatest result of this theory is that the mass superselection rule is not strictly necessary.
Moreover, the ability to describe mass decays and to grant that the mass keeps up with energy changes, such as changes in binding energy, further strengthen the conventional quantum theory.

%%%%%%%%%%%%%%%%%%%%%%%%%%%%%%%%%%%%%%%%%%%%%%%%%%%%%%%%%%%%
\chapter{Variable mass from the Dirac \acl{QW} covariance}\chaptermark{Variable mass from the Dirac \acs{QW} covariance}
%%%%%%%%%%%%%%%%%%%%%%%%%%%%%%%%%%%%%%%%%%%%%%%%%%%%%%%%%%%%
\label{cap:automata}

In this last chapter we show how the upgrade of the particle \emph{mass} from invariant parameter to variable can be a direct consequence of elementary assumptions on the dynamics of a physical system.

Recently in \cite{DAriano2014}, it has been proposed a discrete theory for quantum field dynamics based on finite dimensional quantum systems in interaction.
Assuming the locality, homogeneity, and unitarity of the interaction leads to describe particles evolution via a discrete time \acf{QW}.  Here we point out that the covariance under change of reference frame of the resulting \ac{QW} theory cannot leave the value of the particle mass invariant.

Due to the discreteness of the walk framework the Lorentz covariance is generally broken, with the usual symmetries recovered in the relativistic limit.
The symmetry of the \acl{QW} is instead provided by a non-linear realization of the De Sitter group.

\section{Dirac Quantum Walk in one space dimension}

A (classical) \keyword{random walk} describes a particle which moves in discrete time steps and with certain probabilities from one lattice position to the neighboring lattice positions.
This simple model of discrete evolution had a broad spectrum of applications in several fields of mathematics, physics, chemistry, computer science, natural sciences, and economics.
We are here interested in the quantum version of such a random walk, denoted \keyword{\acl{QW}}, firstly introduced in \cite{Aharonov1993} where the motion (right or left) of a spin-1/2 particle is decided by a measurement of the $z$-component of its spin.
Subsequently, the measurement was replaced by a unitary operator on the internal space, also known as \keyword{coin space}\footnote{For further readings see \cite{Ambainis:2001:OQW:380752.380757}.}, determining the evolution of the internal degree of freedom of the system.
At the first stage, this model gained the interest of the quantum information community due to its capability of providing a computational speedup over classical random walks for a class of relevant problems.
Beside their relevance in computer science, \acsp{QW} provide a natural setting for an elementary quantum theory of non-interacting quantum fields.
Some relevant results are explained in \cite{Strauch2006,Meyer1996,Bialynicki-Birula1994}, in which many physical equations and phenomena are derived from pure mathematical and quantum principles by means of \acsp{QW}.

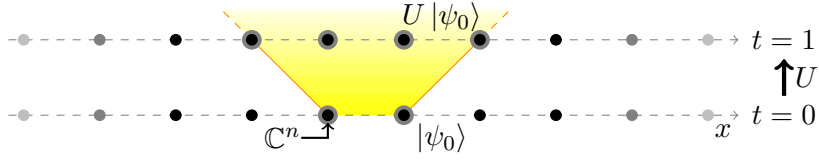
\begin{figure}
	\caption{Schematic representation of a one-dimensional quantum walk.}
	\centering
	\label{pic:QW}
	% !TeX encoding = UTF-8
% !TeX root = ./tesi.tex

% CC-BY Matteo Lugli (matteo.lugli@outlook.com)
% Quest'opera è stata rilasciata con licenza Creative Commons Attribuzione 4.0 Internazionale.
% Per leggere una copia della licenza visita il sito web http://creativecommons.org/licenses/by/4.0/.

% Monodimensional quantum walk

\begin{tikzpicture}
	% Propagazione
	\shade[top color=yellow!10, bottom color=yellow] (4,0) -- (2.6,1.4) -- (6.4,1.4) -- (5,0) -- cycle;
	\draw[orange] (4,0) -- (3,1);
	\draw[orange] (5,0) -- (6,1);
	\draw[orange,dashed] (3,1) -- (2.6,1.4);
	\draw[orange,dashed] (6,1) -- (6.4,1.4);
	
	% Rette di Z
	\draw[gray,dashed,->] (-0.2,0) -- (9.4,0);
	\draw[gray,dashed,->] (-0.2,1) -- (9.4,1);
	
	% Stato iniziale
	\filldraw[gray] (4,0) circle (3.5pt);
	\filldraw[gray] (5,0) circle (3.5pt);
	
	% Stato finale
	\filldraw[gray] (3,1) circle (3.5pt);
	\filldraw[gray] (4,1) circle (3.5pt);
	\filldraw[gray] (5,1) circle (3.5pt);
	\filldraw[gray] (6,1) circle (3.5pt);
	
	% Reticolo
	\filldraw[black!25!white] (0,0) circle (2pt);
	\filldraw[black!50!white] (1,0) circle (2pt);
	\foreach\x in {2,...,7}
		\filldraw[black] (\x,0) circle (2pt);
	\filldraw[black!50!white] (8,0) circle (2pt);
	\filldraw[black!25!white] (9,0) circle (2pt);
	
	\filldraw[black!25!white] (0,1) circle (2pt);
	\filldraw[black!50!white] (1,1) circle (2pt);
	\foreach\x in {2,...,7}
		\filldraw[black] (\x,1) circle (2pt);
	\filldraw[black!50!white] (8,1) circle (2pt);
	\filldraw[black!25!white] (9,1) circle (2pt);
		
	% Tempi
	\node (TempoZero) at (10,0) {$t=0$};
	\node (TempoUno) at (10,1) {$t=1$};
	
	% Evolutore
	\draw[ultra thick,->] (TempoZero.north) -- (TempoUno.south);
	\node () at (10.3,0.5) {$U$};
	
	% Stati
	\node () at (5.5,-.3) {$\ket{\psi_0}$};
	\node () at (5.5,1.3) {$U\ket{\psi_0}$};
	
	% Grado di libertà interno
	\node at (-1,0) {$\mathbb{C}^n \otimes$};
	\node at (-1,1) {$\mathbb{C}^n \otimes$};
	
	% Asse
	\node () at (9.2,-.2) {$x$};
\end{tikzpicture}
\end{figure}

To our purposes we can focus on \acsp{QW} in one space dimensions.
We consider the one dimensional lattice $\mathbb{Z}$ where at each site there is a finite dimensional quantum system represented by the Hilbert space $\mathbb{C}^n$, for some $n \in\mathbb{N}$, as depicted in \cref{pic:QW}.
The total Hilbert space of the system is then $\mathbb{C}^2\otimes\ell_2(\mathbb{Z})$ and we study the \emph{\acl{QW}}, \ie{} the evolution operator $U$, that satisfies the following hypotheses:
\begin{center}
	\begin{tabular*}{\textwidth}{@{}lll@{\extracolsep{\fill}}l@{}}
		$\bullet$	& \textbf{Unitarity}	& Unitarity of the evolution	& $UU^\dagger = U^\dagger U = I$ \\
		$\bullet$	& \textbf{Homogeneity}	& Translational invariance	& $T_n\colon x \mapsto x + n \qcomma{\forall n \in \mathbb{Z}}$ \\
		$\bullet$	& \textbf{Space isotropy}	& Covariance under parity	& $P_x\colon x \mapsto {-x}$ \\
		$\bullet$	& \textbf{Time isotropy}	& Covariance under time reversal	& $P_t\,\colon t \,\mapsto {-t}$ \\
%% Secondo aggiornamento per arXive
		$\bullet$	& \textbf{Locality}	& \multicolumn{2}{l}{$U$ acts only on the two nearest neighbor sites.} \\
	\end{tabular*}
\end{center}

As proved in \cite{Meyer1996}, no \acs{QW} with $n = 1$ satisfies the above hypothesis while solutions exist for internal degree of freedom $n = 2$.
Accordingly, the total Hilbert space is $\mathbb{C}^2\otimes\ell_2(\mathbb{Z})$ for which we will use the factorized orthonormal basis $\ket{s}\ket{x}$, where for $\ket{s}$ we consider the canonical basis corresponding to $s = l,r$.
The \acs{QW} will then evolve the initial state $\ket{\psi}$ from discrete time $t_0 = 0$ to $t = 1$.
In the $\ket{s}\ket{x}$ representation all unitary matrices $U$ can be written as follows
\begin{equation}
	U := A\otimes S + B\otimes S^\dagger + C\otimes I
\end{equation}
where $A,B,C$ are $2\cross2$ transition matrixes and $S$ is the shift operator $S\colon x \mapsto x + 1$, $S^\dagger$ its adjoint\footnote{Or equivalently, $S = T_1$ and $S^\dagger = T_{\minus 1}$ with the previous notation.} and $I$ the identity on $\mathbb{C}^2$.
In Ref. \cite{Bisio2015} it has been shown that all the admissible solutions are unitarily equivalent to the \acs{QW} having the following transition matrixes
\begin{gather}
	A = \left(\begin{matrix}
	n & 0 \\ 
	0 & 0
	\end{matrix} \right), \quad
	B = \left(\begin{matrix}
	0 & 0 \\ 
	0 & n
	\end{matrix} \right) \\
	C = \left(\begin{matrix}
	0 & -im \\ 
	-im & 0
	\end{matrix} \right) \\
\intertext{where}
	\label{eq:massCondition}
	n^2 + m^2 = 1 \qcomma{n,m\in\mathbb{R}} .
\end{gather}
Then, summing up we find
\begin{equation}\label{eq:QW}
	U := \sum_x
	\begin{pmatrix}
		n \ketbra{x-1}{x} &   -im \ketbra{x}{x} \\
		-im \ketbra{x}{x} &   n \ketbra{x+1}{x} 
	\end{pmatrix} \qcomma{n^2 + m^2 = 1}
\end{equation}
describing a \acl{QW} on the Hilbert space $\mathbb{C}^2\otimes\ell_2(\mathbb{Z})$.

Tanks to the translational invariance of $U$, it is convenient to move to the momentum
representation
\begin{equation}\label{vk}
	\ket{\psi_s}\ket{k} :=\frac{1}{\sqrt{2\pi}}  \sum_{x} e^{\minus ikx} \ket{\psi_s}\ket{x} \qcomma{k\in {[{\minus\pi},\pi]}},
\end{equation}
and $U$ becomes
\begin{equation}\label{eq:QWmomentum}
	U = \int_{\minus\pi}^{\pi}\dd{k}U(k)\otimes\ketbra{k}{k}, \quad U(k) = \begin{pmatrix}
		n e^{ik} & -i m \\
		-i m & n e^{\minus ik}
	\end{pmatrix}.
\end{equation}
Notice that discreteness bound momenta to the Brillouin zone ${[{\minus\pi},\pi]}$ as in solid-state theory.
We diagonalize the unitary matrix $U(k)$
\begin{equation}\label{eq:eigenstates}
\begin{split}
	U(k)\ket{s}_k = e^{\minus is \omega(k)}\ket{s}_k, & \qquad \omega(k) = \arccos(n \cos(k)) \\
	\ket{s}_k := \frac{1}{\sqrt{2}}
	\begin{bmatrix}
		\sqrt{1-sv(k)}\\
		s\sqrt{1+sv(k)}
	\end{bmatrix}, & \qquad s=\pm,\; v(k) := \pdv{\omega}{k} (k) .
\end{split}
\end{equation}
We get the function $\omega(k)$ by using the invariance of the trace under change of basis:
\begin{align}
	\Tr\Lambda &= \Tr U \\
\intertext{where $\Lambda$ is the diagonal form of $U(k)$. Thus we have}
	e^{is \omega(k)} + e^{\minus is \omega(k)} &= n\left(e^{ik} + e^{\minus ik}\right) \\
	\label{eq:dispersionStep3}
	\cos(\omega(k)) &= n\cos(k) \\
\intertext{and finally}
	\omega(k) &= \arccos(n\cos(k))
\end{align}
as in \cref{eq:eigenstates}.

It is easy to evaluate the logarithm of $U$ ($H|e^{\minus i H}:= U$) as follows
\begin{equation}\label{eq:hamiltonian}
\begin{split}
	H = \int_{\minus\pi}^{\pi}\dd{k}H(k) & \otimes\ketbra{k}{k}, \\
	H(k) & = \omega(k)\left(\ket{+}_k\bra{+}_k - \ket{-}_k\bra{-}_k\right) \\
	& = \frac{\omega(k)}{\sin(\omega(k))}(-n\sin(k) \sigma_3+m\sigma_1)
\end{split}
\end{equation}
where $\sigma_i,\; i=1,2,3$ denote the usual Pauli matrices. 

The function $\omega(k)$ is the dispersion relation of the walk, which recovers the usual Dirac one $\omega(k) = \sqrt{k^2 + m^2}$ in the limit $k,m \ll 1$ and $k/m \gg 1$.
The $s = +1$ eigenvalues correspond to positive-energy particle states, whereas the negative $s = -1$ eigenvalues correspond to negative-energy anti-particle states. Notice that the operator $H$ regarded as an Hamiltonian would interpolate the evolution to continuous time as $U(t)\equiv U^t$.

In \cite{Bisio2016,Bibeau-Delisle2015}, the covariance of the above walk has been studied.
One can immediately notice that, if we consider the usual linear representation of the Lorentz group $\mathsf{SO}^+(1,1)$
\begin{equation}
	L_\beta\colon \left(\omega,k\right) \mapsto \left(\omega',k'\right) = \gamma\left(\omega - \beta k,k - \beta\omega\right)
\end{equation}
where $\gamma = \frac{1}{\sqrt{1 - \beta^2}}$ is the Lorentz factor, the dispersion relation $\omega(k)$ of \cref{eq:eigenstates} is not invariant under it.
In order to preserve the Lorentz group structure, non-linear representations are instead taken into account.
The study of all of them is very hard and we will address exclusively those that grant the invariance of the dispersion relation $\omega(k)$ only, which is of course a superset.

By elevating to the second power \cref{eq:massCondition,eq:dispersionStep3} we find:
\begin{align}
	\cos[2](\omega(k)) &= n^2 \cos[2](k) = \left(1 - m^2\right)\cos[2](k) \\
	1 - \sin[2](\omega(k)) &= - \left(1 - m^2\right)\sin[2](k) + 1 - m^2
\end{align}
and hence
\begin{equation}\label{eq:dispersionStep6}
	\sin[2](\omega(k)) - \left(1 - m^2\right)\sin[2](k) - m^2 = 0
\end{equation}
which is \cite[eq.~(19)]{Bisio2016} in the monodimensional case.
In the same article, it is remarked that the covariance under change of reference cannot leave the value of $m$ invariant as one can prove with ease from \cref{eq:dispersionStep6}.
Hence, we are forced to expand the usual Lorentz symmetry (or more generally the Poincaré one) in a similar way to what have already done in \cref{cap:theory}.
The symmetry group for the unidimensional \acs{QW} requires a further variable and is indeed the de Sitter $\mathsf{SO}(1,2)$ group ($\mathsf{SO}(1,4)$ in three dimensions).

The common physical interpretation of \acsp{QW} is the evolution of a ½-spin mass field.
On the one hand, if we consider the parametrization of \cref{eq:QWmomentum}, we notice that for $m = 0$ we have $C = 0$ or
\begin{equation}
	U := \sum_x
	\begin{pmatrix}
	\ketbra{x-1}{x}	& 0 \\
	0				& \ketbra{x+1}{x} 
	\end{pmatrix}
\end{equation}
and hence, the state of a site is forced to move to the maximum speed $c = 1$, \ie{} one site per step.
On the other hand, for both $m = \pm1$ we have $A, B = 0$ or
\begin{equation}
	U := \sum_x
	\begin{pmatrix}
	0					& \mp i\ketbra{x}{x} \\
	\mp i\ketbra{x}{x}	& 0 
	\end{pmatrix}
\end{equation}
and the state localized on a single site for each site does not move.
From these two limit cases, one is induced to interpret the parameter $m$ of the \acl{QW} as the mass of the field.
A rigorous comparing between the \acs{QW} \cref{eq:QW} evolution and the usual Dirac dynamics, which ultimately provides the
physical interpretation of the walk parameter $m$, $k$ and $w$ as the Dirac field mass, momentum and energy respectively, has been presented in \cite{Bisio2015}.

If we apply the hypothesis of \Greenberger{} to \aclp{QW}, \ie{} mass and proper time are conjugated variables, we observe an interesting point.
Since mass $m$ is bound into the interval $m \in {[{\minus 1},1]}$ from \cref{eq:massCondition} and the dynamics are exactly the same for the limit cases $m = \pm1$, the parameter $m$ lies actually on a 1-torus.
The Fourier transform of a variable defined on a circle is discrete, and therefore proper time would be discrete, consistently with our premise of integer global time $t$.
This is barely a consistent speculation and no more; there is still a huge work to understand what are the consequences of our interpretations.
Nevertheless, \acsp{QW} offer a promising groundwork to deal with
foundational aspects of quantum field theory.
The strengths of the framework are the lack of quantization rules, since the systems at play are quantum \emph{ab initio}, and the chance of thoroughly analyze the consequences of any further assumption on the dynamics of the model, keeping the assumptions as elementary as possible.

%%%%%%%%%%%%%%%%%%%%%%%%%%%%%%%%%%%%%%%%%%%%%%%%%%%%%%%%%%%%
% Inizio appendici
%%%%%%%%%%%%%%%%%%%%%%%%%%%%%%%%%%%%%%%%%%%%%%%%%%%%%%%%%%%%
\appendix
%%%%%%%%%%%%%%%%%%%%%%%%%%%%%%%%%%%%%%%%%%%%%%%%%%%%%%%%%%%%
\chapter{Gauge condition for the gravitational potential}
\chaptermark{Gauge condition}
%%%%%%%%%%%%%%%%%%%%%%%%%%%%%%%%%%%%%%%%%%%%%%%%%%%%%%%%%%%%
\label{app:gaugeCondition}

As already stated in \cref{subsec:classicalGrav}, since our classical theory of particles with variable mass is the non-relativist limit of a gauge field one elaborated in \cite{Greenberger1970-I}, we must be able to solve a non-linear gauge condition.
All gravitational potentials $\varphi$ must satisfy:
\begin{equation}
	\pdv{\varphi}{t} + \frac{g}{c^2}\varphi\pdv{\varphi}{\tau} = 0 . \tag{\ref{eq:gaugeCondition}}
\end{equation}
where $g$ is the coupling constant.
To simplify notation, in the following calculation we will consider the dimensionless potential:
\begin{equation}\label{eq:gaugeSubstitution}
	\psi = \frac{g}{c^2}\varphi
\end{equation}
so that \cref{eq:gaugeCondition} becomes:
\begin{equation}\label{eq:gaugeConditionSimplified}
	\pdv{\psi}{t} + \psi\pdv{\psi}{\tau} = 0 .
\end{equation}
Thus, we are seeking a potential of the form $\psi = \psi(x_i,\tau,t)$ which satisfies \cref{eq:gaugeConditionSimplified}.
The condition of \cref{eq:gaugeCondition} becomes linear by inverting $\tau = \tau\left(x_i,t,\psi\right)$ and considering that:
\begin{align}
	\psi\left(x_i,\tau,t\right) &= - \pdv{\psi}{t} / \pdv{\psi}{\tau} = \pdv{\tau}{t} , \\
\intertext{if we integrate with respect to $t$ we come to the general solution}
	\label{eq:gaugeConditionSolution}
	\tau\left(x_i,t,\psi\right) &= t\psi + f\left(x_i,\psi\right)
\end{align}	
which is equivalent to
\begin{equation}
	f(x_i,\varphi,\tau - t\psi) = 0 .
\end{equation}

In order to solve the gauge condition for \cref{subsec:decay}, we consider the proper time Hamilton's equation for the case:
\begin{equation}\label{eq:decayGaugeSolutionStep1}
	\dot{\tau} = 1 + \frac{\varphi}{c^2} = 1 + \frac{\psi}{g}.
\end{equation}
If we differentiate over proper time $\tau$ \cref{eq:gaugeConditionSolution} we have:
\begin{equation}\label{eq:decayGaugeSolutionStep2}
	1 = t\pdv{\psi}{\tau} + \pdv{f}{\psi}\pdv{\psi}{\tau} = \left(t + \pdv{f}{\psi}\right)\pdv{\psi}{\tau}
\end{equation}
whereas by totally differentiating over time $t$ and substituting with \cref{eq:decayGaugeSolutionStep2} we come to:
\begin{equation}\label{eq:decayGaugeSolutionStep3}
	\dot{\tau} = \psi + t\dot{\psi} + \pdv{f}{\psi}\dot{\psi} = \psi + \left(t + \pdv{f}{\psi}\right)\dot{\psi} = \psi + \dot{\psi}/\pdv{\psi}{\tau}
\end{equation}
where $\pdv*{f}{x_i} = 0$ as the particle is at rest and therefore $f$ is independent of $x_i$.
By combining \cref{eq:decayGaugeSolutionStep1,eq:decayGaugeSolutionStep2,eq:decayGaugeSolutionStep3} we obtain the result:
\begin{equation}\label{eq:gaugeConditionResult}
	\dot{\psi} = \pdv{\psi}{\tau} \quad\longleftrightarrow\quad \dot{\varphi} = \pdv{\varphi}{\tau}
\end{equation}
recalling the substitution of \cref{eq:gaugeSubstitution}.

The case of \cref{subsec:bindingEnergy} is slightly different.
For proper time, as in \cref{eq:HamiltonEquationPhiV}, we have:
\begin{align}
	\label{eq:bindingHamiltonEq}
	\dot{\tau} &= 1 - \dfrac{1}{2}\frac{\vb{v}^2}{c^2} + \frac{\varphi}{c^2} = 1 - \dfrac{1}{2}\frac{\vb{v}^2}{c^2} + \frac{\psi}{g},
\intertext{if we integrate along the particle trajectory $\gamma$ we come to}
	\label{eq:bindingGaugeSolutionStep1}
	\tau &= t\frac{\psi}{g} - \frac{1}{2}\int_{\gamma}\frac{\vb{v}}{c^2}\vdot\vb{\dd{x}} + f\left(\psi\right).
\end{align}
Now, we evaluate the derivative of \cref{eq:bindingGaugeSolutionStep1} with respect to proper time $\tau$
\begin{align}
	\label{eq:bindingGaugeSolutionStep2}
	1 = \frac{t}{g}\pdv{\psi}{\tau} + \pdv{f}{\psi}\pdv{\psi}{\tau} &\longrightarrow \left(\frac{t}{g} + \pdv{f}{\psi}\right)\pdv{\psi}{\tau} = 1
\intertext{and to position $x_i$}
	0 = \frac{t}{g}\pdv{\psi}{x_i} - \frac{1}{2}\frac{\dot{x}_i}{c^2} + \pdv{f}{\psi}\pdv{\psi}{x_i} &\longrightarrow \left(\frac{t}{g} + \pdv{f}{\psi}\right)\pdv{\psi}{x_i} = \frac{1}{2}\frac{\dot{x}_i}{c^2}
\end{align}
so we have
\begin{equation}\label{eq:bindingGaugeResult}
	\pdv{\psi}{x_i} = \frac{1}{2} \frac{\dot{x}_i}{c^2}\pdv{\psi}{\tau} \quad\longleftrightarrow\quad \pdv{\varphi}{x_i} = \frac{1}{2} \frac{\dot{x}_i}{c^2}\pdv{\varphi}{\tau} .
\end{equation}
because of \cref{eq:gaugeSubstitution}.
The total time derivative of \cref{eq:bindingGaugeSolutionStep1} is
\begin{align}
	\dot{\tau} &= \frac{\psi}{g} + t\frac{\dot{\psi}}{g} - \frac{1}{2}\frac{\vb{v}^2}{c^2} + \pdv{f}{\psi}\dot{\psi} = \frac{\psi}{g} - \frac{1}{2}\frac{\vb{v}^2}{c^2} + \left(\frac{t}{g} + \pdv{f}{\psi}\right)\dot{\psi}
\intertext{from \cref{eq:bindingGaugeSolutionStep2,eq:bindingHamiltonEq} we have}
	\dot{\tau} &= \frac{\psi}{g} - \frac{1}{2}\frac{\vb{v}^2}{c^2} + \dot{\psi}/\pdv{\psi}{\tau} = 1 - \dfrac{1}{2}\frac{\vb{v}^2}{c^2} + \frac{\psi}{g} 
\end{align}
Finally, we have again
\begin{equation}
\dot{\psi} = \pdv{\psi}{\tau} \quad\longleftrightarrow\quad \dot{\varphi} = \pdv{\varphi}{\tau} \tag{\ref{eq:gaugeConditionResult}} .
\end{equation}

\backmatter
%%%%%%%%%%%%%%%%%%%%%%%%%%%%%%%%%%%%%%%%%%%%%%%%%%%%%%%%%%%%
% Bibliografia
%%%%%%%%%%%%%%%%%%%%%%%%%%%%%%%%%%%%%%%%%%%%%%%%%%%%%%%%%%%%
\bibliography{bib/Greenberger,bib/automata,bib/articles,bib/books}
\bibliographystyle{halpha}

%%%%%%%%%%%%%%%%%%%%%%%%%%%%%%%%%%%%%%%%%%%%%%%%%%%%%%%%%%%
% Indice analitico
%%%%%%%%%%%%%%%%%%%%%%%%%%%%%%%%%%%%%%%%%%%%%%%%%%%%%%%%%%%
\cleardoublepage
\printindex

\end{document}